\documentstyle[12pt]{article}
\renewcommand{\baselinestretch}{1.8}
\newcounter{appx}
\newcommand{\appeqn}{\stepcounter{appx}\setcounter{equation}{0}%
\renewcommand{\theequation}{\Alph{appx}\arabic{equation}}}
\begin{document}
\title{\sc Totally Chaotic Poissonian-Like Sources in Multiparticle
Production Processes?}
\vspace{-5mm}
\author{M.Biyajima$^{1}$\thanks{e-mail: minoru44@jpnyitp.bitnet},
N.Suzuki$^{2}$, G.Wilk$^{3}$\thanks{e-mail: wilk@fuw.edu.pl}
and Z.W\l odarczyk$^{4}$\thanks{e-mail: wsp-fiz@srv1.tu.kielce.pl}}
\date{}
\maketitle
\vspace{-8mm}
\begin{center}
$^1${\it Department of Physics, Faculty of Sciences,\\
\vspace{-2mm}
Shinshu University,  Matsumoto 390, Japan}\\
$^2${\it Matsusho Gauken Junior College, Matsumoto 390-12, Japan}\\
$^3${\it Soltan Institute for Nuclear Studies, Warsaw, Poland,}\\
$^4${\it Institute of Physics, Pedagogical University, Kielce, Poland}
\end{center}
\begin{abstract}
In all multiparticle processes the concept of sources directly emitting
finally observed secondaries (mostly pions) plays crucial role. Here we
shall present them from yet another point of view in which elementary
sources composing all processes (from $e^+e^-$ annihilation, via $pp$
up to $AA$ interactions) remain both totally 'chaotic' and Poissonian
at the same time. \\

\end{abstract}

\newpage

\section{Introduction}

For time being the multiparticle production processes can be
described only in a phenomenological way. As a rule they are visualised
as proceeding in two steps: first a number of more or less defined
intermediate objects (we shall called them {\it sources}) is formed and
next follows their hadronization. Sources are heavy, nonresonant and
unstable objects and models differ in their definition and in details
of their hadronization \cite{MOD,B}. \\

The notion of sources is essential in explaining such general features
of multiparticle production processes as the broadening of the multiplicity
distributions $P(n)$ when the complexity of the colliding objects
increases, cf. Fig. 1, and that the second factorial cumulant is almost
constant (it decreases only insignificantly, cf. Fig. 2). \\

There is another possible characteristic of the source discussed
recently \cite{CD}. It is represented by parameter $\lambda$ defined as
\begin{equation}
\lambda = \left[ C_2(p_1,p_2) - 1 \right]_
          {{\rm lim} |p_1-p_2|\rightarrow 0} , \label{eq:L}
\end{equation}
where $C_2(p_1,p_2)$ is the Bose-Einstein correlation (BEC) term of two
identical bosons with momenta $p_1$ and $p_2$ \cite{LOR}. This
parameter was introduced as a measure of {\it correlation strength} in
the simplest possible one-dimensional parametrization of $C_2$ in order
to reduce systematic errrors when fitting the experimental results with
theoretical curves (cf., for example, \cite{DEUT}). It has got very
quickly quantum-optical interpretation as the {\it chaoticity
parameter} \cite{WEINER,GKW} and is widely accepted under this name in
majority of phenomenological approaches to BEC. In this interpretation
$\lambda =1$ signals totally chaotic (usually understood as thermal)
emission of secondaries whereas $\lambda =0$ means that they are
radiated in a laser-like fashion. In other words, one expects that
quantum-mechanical phases in different space-time points of the
hadronization region are totally uncorrelated in the former case and
fixed by one value in the later one \cite{GKW,CD}.\\

It was known from the very beginning that such interpretation of
$\lambda $ has its severe limitations and that it is also affected by
the type of the parametrization of BEC used \cite{QQ1,BOWA,DEUT}.
Especially emabarasing in this respect, although not much pursued
(cf. \cite{CD}), are observations that:
\begin{itemize}
\item in in $e^+e^-$ annihilation processes $\lambda$ is nearly maximal
\cite{BMW} and practically $\lambda$ does not depend on the
multiplicity of produced secondaries;
\item in the more complex $NN$ collisions $\lambda$ drops considerably
and decreases with increasing multiplicity \cite{F1};
\item this trend seems to continue when proceeding to $hA$ collisions
where $\lambda_{hA} < \lambda_{hh}$ is apparently observed \cite{CAMP1}
and in $AA$ collisions where $\lambda $ decreases with the atomic
masses $A$ \cite{CAMP2}.
\end{itemize}

\section{Totally chaotic  elementary emitting cels in the simplest
$e^+e^-$ collisions}

Taken naively these observations could indicate an increase of {\it
coherence} from $e^+e^-$ to $AA$ collisions, an impossible conclusion
from the point of view of the observed in Fig. 1 behaviour of $P(n)$.
In \cite{CD} a discussion of $\lambda $ trying to reconcile experimental
observations presented above was given with elementary source being
totally {\it coherent}. We shall demonstrate now that one can obtain
equally good description of data with {\it totally chaotic} elementary
emitting cells {\it EEC} and observed changes of parameter $\lambda $
(defined in eq.(\ref{eq:L})) being caused by the increasing complexity
of reaction - increasing and fluctuating number of such cells. \\

Our {\it elementary emitting cell} ({\it EEC}) is assumed to produce bosons
(mainly pions) in only one momentum state $|p\rangle $. Therefore, due
to the Bose-Einstein statistics, the multiplicity distribution of pions
from such source is purely of geometric (Bose-Einstein) type
corresponding to the {\it chaotic} field limit
\begin{equation}
P(n) = \frac{1}{1 + \langle n \rangle } \left( \frac{\langle n \rangle }{1
+ \langle n \rangle } \right)^n   \label{BE}
\end{equation}
and, correspondingly, for a single {\it EEC} one has $\lambda =1$ (i.e., all
particles are fully correlated in the sense that presence of any one of
them stimulates additional emission of other secondaries which is
limited only by the energy-momentum conservation constrains not
considered here). On the contrary, particles originating from different
{\it EEC}'s are totally uncorrelated (again, in the above mentioned
sense), therefore for them $\lambda = 0$. Notice that our {\it EEC}'s
are more fundamental than sources mentioned at the beginning
(fireballs, clusters, strings etc). Actually these sources, which we
shall denote {\bf S} in what follows, contain always a whole spectrum
of {\it EEC}'s corresponding to the distribution of momenta of the
produced particles ( $\sim \exp(-p/p_0)$).  We shall assume here
that in multiparticle production processes all {\it EES}'s are produced
independently. In the case when we have $n$ originating from $k$
{\it EEC}'s we are immediately lead to the famous negative binomial
(NB) distribution \cite{B,CC}
\begin{eqnarray}
P(n;k) &=& \sum_{n_1,n_2,\dots,n_k} \prod_i \left[
\frac{\left( \frac{\langle n\rangle}{k}\right) ^{n_i}}
     {\left( 1 + \frac{\langle n\rangle}{k}\right) ^{n_i + 1}}\right]
     \nonumber \\
       &=& { n + k -1 \choose  n}
       \frac{\left( \frac{\langle n\rangle}{k}\right)^n}
       {\left( 1 + \frac{\langle n\rangle}{k}\right) ^{n + k} }
       \qquad \stackrel{k\to \infty }{\longrightarrow} \qquad
       P(n) = \frac{\langle n \rangle ^n}{n!} {\rm e} ^{-\langle n\rangle}
          \label{eq:NB}
\end{eqnarray}
which in the case of large number of {\it EEC}'s (understood as
$\langle n\rangle /k << 1$) leads to Poissonian distribution of
produced secondaries $n$. For widely tested experimentally variance of
multiplicity distribution it means that
\begin{equation}
D^2 = \langle n\rangle \left( 1 + \frac{\langle n \rangle}{k} \right)
       \qquad \stackrel{k\to \infty }{\longrightarrow} \qquad
       \langle n \rangle .  \label{eq:VAR}
\end{equation}
Therefore, it is quite natural in our picture that in the $e^+e^-$
annihilation processes we shall indeed at the same time observe both
$P(n)$ being poissonian(-like) and $\lambda$ near $1$. \\

\section{More complex hadronic and nuclear collisions}

When one proceeds to more complex hadronic collisions the first thing
to recognize is that now the broad hadronic multiplicity distributions
can arise in a natural way from any incoherent superposition of many
narrow ('$e^+e^-$-type like') sources {\bf S} \cite{BAR}. Suppose that we
consider a fluctuating number $C$ of such independent narrow
(Poissonian) {\bf S}'s with $n_i$ particles in each and with total
multiplicity $N = \sum^C_{i=1} n_i$ and mean $\langle N \rangle =
\langle n \rangle \langle C \rangle $. As a result we have then a
compound Poissonian distribution \cite{SR}
\begin{equation}
P(N) = \sum ^{\infty }_{C=0} \frac{\left( C \langle n \rangle \right) ^N}
        {N!}\cdot {\rm e}^{-C\langle n \rangle}\cdot
        \frac{\langle C\rangle ^C {\rm e}^{- \langle C\rangle}}{C!}
        \label{eq:CP}
\end{equation}
with the same variance as in the case of NB type of $P(n)$ (cf.
eq.(\ref{eq:VAR}) with $\langle C\rangle $ replacing $k$)
\cite{F2}. \\

Proceeding now to the most complex nuclear interactions one finds
that they are dominated almost exclusively by the geometry of
collision. The characteristic broad and flat shape of nuclear
multiplicity distribution $P(N)=\sum_{\nu}p(\nu)P(N|\nu)$ emerges from
the smooth behaviour of distribution of the number of participating
nucleons $\nu,\quad p(\nu)\sim b(\nu)\frac{db}{d\nu}$, over impact
parameter $b$ in a wide region of $\nu$ \cite{ZL}. In general one has
\begin{equation}
\frac{D^2(N)}{\langle N\rangle ^2} = \frac{D^2(\nu)}{\langle
\nu\rangle^2} + \frac{1}{\langle \nu\rangle}\cdot \frac{D^2(n)}{\langle
n \rangle ^2}  \label{eq:NUC2}
\end{equation}
(where $n$ denotes multiplicity in hadronic collisions) with the first
term being completely due to the nuclear geometry
and dominating in the midle part of $\nu$ for minimum-bias events where
$p(\nu)\sim $ const. In this region one gets multiplicity
moments independent of the target mass and given by simple formula:
\begin{equation}
\mu _q = \frac{\langle \nu ^q\rangle}{\langle \nu \rangle ^q } =
\frac{2^q}{q+1}. \label{eq:NUCMOM}
\end{equation}

This result holds for any situation where one has a number $C$ of type
{\bf S} sources decaying independently into $n_i$ particles each
($i=1,\dots,C$). Namely, the normalized factorial moments $F_q$ of the
total distribution and those for each source {\bf S}, $F^{(S)}_q$, can be
related to each other \cite{CCS}, for example:
\begin{equation}
F_2 = F^S_2 + \frac{F^{(S)}_2}{\langle C\rangle}  \label{eq:FF}
\end{equation}
with $F^S_2 = \sum_C C(C - 1) P(C)/\langle C\rangle ^2$ being the
normalized factorial moment of the distribution of {\bf S}'s and
$\langle C\rangle $ their mean number whereas $F^{(S)}_2=\langle
n(n-1)\rangle/\langle n\rangle ^2$. Fig. 2 shows this quantity
calculated as an average over $10^3$ bins in the phase space as a
function of the mean number of {\bf S}'s for different types of their
distributions. Notice that $F_2$ is by definition the same for small
$\langle C\rangle $ and that different types of fluctuations result in
different dependence of $F_2$ on $\langle C\rangle$. Notice also that
starting from the value $F_2 = 1.5$ for single source (corresponding,
for example, to "elementary" $\mu p$ collisions for which $F_2\simeq
1.5$ \cite{MUP} ) $F_2$ decreases very slowly for the flat $P(C)$
distribution tending to the plateau for moderately large number of {\it
S}'s and reaching value of $F_2\simeq 1.3 - 1.35$ which coincides with
that observed in heavy ion data \cite{NA35}. In this way we can
describe this two widely different (from the point of view of their
compositeness) processes \underline{without} necessity of questioning
the independent collision picture in the later as has been advocated
recently in \cite{F}. \\

\section{Summary and conclusions}

As we have just demonstrated our sources {\bf S}'s fit nicely into
description of all type of collisions. Let us return then back to
discussion of the parameter $\lambda$ as defined by eq.(\ref{eq:L}). We
want to show now how it can serve as a measure of sources {\bf S}'s
(instead of being the measure of {\it chaoticity}, the notion of which
does not appear here at all). Let us consider $C$ sources {\bf S}, each
producing (on average) $\langle n\rangle$ like-sign bosons. We have
\begin{equation}
n^{(S)}_{pairs} = \frac{1}{2}\langle n(n-1)\rangle \nonumber
\end{equation}
pairs of such bosons from a single source and, respectively, $\langle
C\rangle$ times this emerging from $\langle C\rangle$ sources:
\begin{equation}
n^S_{pairs} = \langle C\rangle n^{(S)}_{pairs} =
                     \frac{1}{2} \langle C\rangle \langle n(n-1)\rangle
                     \label{eq:NSpairs}
\end{equation}
whereas the \underline{total} number of pairs of like-sign bosons
producing by $\langle C\rangle$ sources is instead equal to (cf.
Appendix A)
\begin{equation}
N^{tot}_{pairs} = \frac{1}{2}\langle C(C-1)\rangle
                                  \langle n\rangle^2 +
                                  \frac{1}{2}\langle C\rangle \langle
                                  n(n-1) \rangle . \label{eq:NTpairs}
\end{equation}
The $\lambda$ parameter is, of course (cf. Appendix B), given by the
ratio of both, which can be write as
\vspace{2mm}
\begin{equation}
\lambda = \frac{\mu_2(n)\langle n\rangle - 1}
          {\left[ \mu_2(C)\langle C\rangle + \mu_2(n) -1 \right]
                                            \langle n\rangle - 1}
        = \frac{\mu_2(n)\langle N\rangle - \langle C\rangle}
               {\langle C\rangle \left[ \mu_2(N)\langle N\rangle -
               1\right] },                   \label{eq:LL}
\end{equation}
\vspace{3mm}
\noindent
where $\langle N\rangle = \langle C\rangle \langle n\rangle $ and
$\mu_2(k)=\frac{\langle k^2\rangle}{\langle k\rangle ^2}$, i.e., it
is the second scaled moment for $k=n,~N$ and $C$, respectively. Note
that, as a consequence of our definition of {\it EEC}, $\lambda =1$ for
single source ($C=1$) and decreases with the number of {\bf S}'s.\\

One can summarize now experimental situation as discussed previously in
terms of our sources {\bf S}.
\begin{itemize}
\item[$(i)$] In $e^+e^-$ annihilations only $1-2$ sources {\bf S} are
produced \cite{AG}. This leads to large values of $\lambda$ and increasing
multiplicity comes mainly from the increasing number of secondaries
from {\bf S}.
\item[$(ii)$] In $pp$ collisions more sources are
produced and multiplicity grows at first with the increasing number of
sources and later, when this number saturates, because of increasing
number of secondaries from {\bf S}, as in $(i)$. This leads to
$\lambda$ decreasing with multiplicity, cf. Fig. 3. Note that the
number of {\bf S}'s, $\langle C\rangle$ increases linearly with $dN/dy$
(with $\sim 2-4$ charged pions per {\bf S}).
\item[$(iii)$] For nuclear collisions the number of the internucleonic
collisions increases with the mass number of colliding nuclei what
results in $\lambda$ decreasing for heavier nuclei, cf. Fig. 4. (Also
here the number of sources {\bf S} deduced from $\lambda$ increases
linearly, this time with the number of participating nucleons,
which is proportional to $A^{1/3}_{\rm T}$).
\end{itemize}

Finally, in Fig. 5 we show analysis of the p+Em collision events
(provided by the IGM event generator \cite{IGM}). Similarly looking
experimental data were used in \cite{FFW} to demonstrate the apparent
increase of the 'coherence' when going in rapidity from target towards
projectile fragmentation regions. This conclusion was based on the
increasing poissonianity of the respective multiplicity distributions
in selected rapidity bins. However, in our case we are getting
precisely the same pattern without invoking any notion of 'coherence'
at all. It is enough that particles produced near the kinematic limits
of reaction originate practically from one source only in which case in
our approach they should be both poisson-like distributed and show
$\lambda =1$.\\

This last observation has some profound consequences which most
probably can be tested in cosmic ray emulsion chambers experiments.
Namely, if (based on the quantum-optical concepts) interpretation of
\cite{FFW} is correct, one should not observe any BEC effects in the
fragmentation  regions  of reactions. However, as demonstrated in
\cite{ZW}, the presence of such correlation (with their full strength,
i.e., with $\lambda = 1$) would explain in a natural and consistent way
many apparently 'strange' effects observed in the mentioned above
cosmic ray experiments (which by definition measure almost exclusively
the fragmentation region of hadronic collisions on air nuclei). In
addition we would predict that, contrary to \cite{FFW}, $\lambda$ will
decrease towards the central region of reaction.\\

\noindent
\underline{Acknowledgements}:~~~
M.B. is grateful for partial support provided by Japanese Grant-in-Aid
for Scientific Research from the Ministry of Education, Science and
Culture (\#. 06640383). N.S. is partially supported by Matsumoto Gauken
Junior College in 1995.\\

\section*{Appendix A}
\appeqn

We shall provide here an elementary derivation of eq.(\ref{eq:LL}). Let
$P_S(C)$ denote the number distribution of sources {\bf S} and $p_1(n)$ the
multiplicity distribution of like-sign particles from a single ($C=1$)
source. Because there are no BEC between particles emitted from different
sources if their phases are random (which we assume here), therefore the
total multiplicity distribution (again of like-sign particles) $P(N)$
is given by
\begin{equation}
P(N) = \sum_C\, P_S(C)\, \sum_{n_1+\cdots+n_C=N} \prod_{i=1}^C\,
       p_1(n_i). \label{eq:Ptot}
\end{equation}
Using generating functions:
\begin{eqnarray}
\pi(z)    &=& \sum_N\, P(N)\, z^N ,\nonumber \\
\pi_S(z)  &=& \sum_C\, P_S(C)\, z^C, \nonumber \\
g(z)      &=& \sum_n\, p_1(n)\, z^n \nonumber
\end{eqnarray}
we have (with $u=g(z)$)
\begin{eqnarray}
\pi(z)  &=& \sum_C\, P_S(C)\, \left[ g(z)\right]^C\, \, =\, \,
            \pi_S\left[ g(z)\right], \nonumber \\
\pi'(z) &=& \pi_S'(u)\, g'(z), \nonumber \\
\pi''(z) &=& \pi_S''(u)\, \left[ g'(z)\right] ^2\, +\, \pi_S'(u)\, g''(z).
           \nonumber
\end{eqnarray}
Because, by definition,
\begin{eqnarray}
\langle n\rangle &=& g'(z)|_{z=1}, \nonumber \\
\langle N(N-1)\rangle &=&  \pi''(z)|_{z=1}, \nonumber \\
\langle C(C-1)\rangle &=&  \pi_S''(z)|_{z=1}, \nonumber \\
\langle n(n-1)\rangle &=&  g''(z)|_{z=1}, \nonumber
\end{eqnarray}
then we get finally that
\begin{equation}
\langle N(N-1)\rangle = \langle C(C-1)\rangle \langle n\rangle^2 +
                        \langle C\rangle \langle n(n-1)\rangle
\end{equation}
from which eq.(\ref{eq:LL}) follows immediately.

\section*{Appendix B}
\appeqn

We shall justify here the eq.(\ref{eq:LL}) in more detail. Let us
introduce (in one dimensional approximation) the usual (cf. refs.
\cite{CD}-\cite{GKW}, especially \cite{QQ1})
two-particle {\it correlation factor}
\begin{equation}
W = 1 + \langle \cos (qr) \rangle  \label{eq:W}
\end{equation}
where $q=p_1 - p_2$ (cf. eq.(\ref{eq:L})) and $\langle \cos (qr)
\rangle $ has symbolic meaning only and can be replaced by any other
suitable expression (like $\langle \exp(-qr) \rangle $) without
changing the outcome of our discussion. The essential point in our
approach consists now in the observation that for $n^{S}_{pairs}$ of
like-sign bosons from {\bf S}-type sources we have such correlation
factor $W$ whereas for the remaining $n_{pairs}^b =
N_{pairs}^{tot}-n_{pairs}^{S}$
(cf. eqs.(\ref{eq:NSpairs},\ref{eq:NTpairs})) it does not appear (i.e.,
the corresponding pairs do not correlate among themselves). On the
other hand, the {\it correlation function} for $N$ particles
\begin{equation}
C_2 = 1 + \lambda \langle \cos (qr) \rangle  \label{eq:C2}
\end{equation}
which already contains parameter $\lambda$ as defined in
eq.(\ref{eq:L}) can be expressed in the following form,
\begin{equation}
C_2 = \frac{n_{pairs}^b + n^S_{pairs}\cdot W}{n_{pairs}^b + n^S_{pairs}}
    = 1 + \frac{n^S_{pairs}}{N^{tot}_{pairs}}\cdot
            \langle \cos (qr) \rangle, \label{eq:C22}
\end{equation}
which immediately provides $\lambda$ in terms of the ratio of pairs of
like-sign bosons leading therefore to eq.({\ref{eq:LL}).

\newpage
\noindent
{\bf Figure Captions}\\

\begin{itemize}
\item[{\bf Fig. 1}] Comparison of: $(a)$ - shapes of multiplicity
distributions $\psi (z=\frac{N}{\langle N\rangle })=\langle N \rangle
P(N)$ and $(b)$ - the energy dependence of the ratio of the dispersion
to the average multiplicity, $\frac{D}{\langle N\rangle}$, for $e^+e^-$
\cite{ee}, $pp$ \cite{pp} and $AA$ \cite{AA} data.
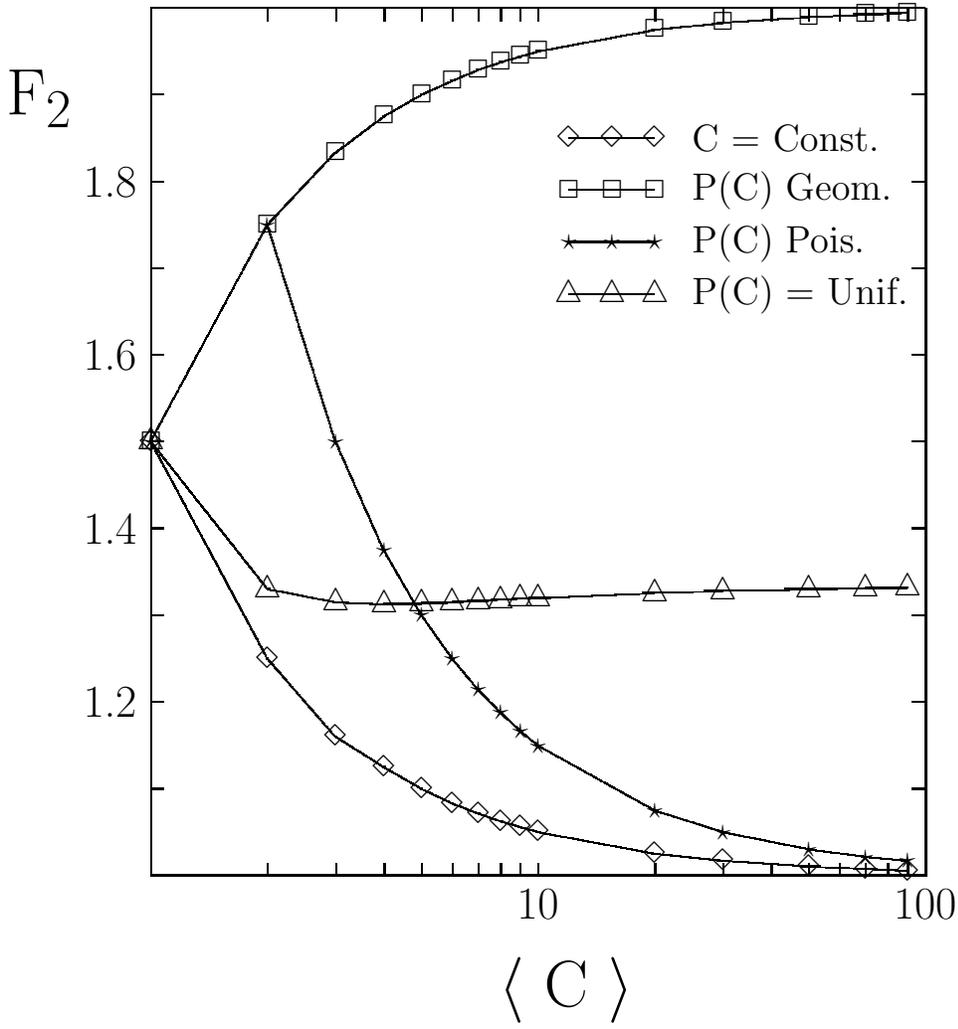
\item[{\bf Fig. 2}] Second normalized factorial moment $F_2$
for $10^3$ bins as function of number $\langle C\rangle$ of sources
{\bf S}  for different types of their distribution.
\item[{\bf Fig. 3}] $(a)$ - Parameter $\lambda$ (as given by
eq.(\ref{eq:L})) as function of multiplicity per unit rapidity,
$dN/dy$; $(b)$ - the same with $\lambda$ replaced by the  mean number
of sources {\bf S}, $\langle C\rangle$ obtained from eq.(\ref{eq:LL}).
Data are from \cite{DATA}.
\item[{\bf Fig. 4}] $(a)$ - Parameter $\lambda$ (cf. eq. (\ref{eq:L}))
as function of the atomic number of target nucleus, $A_T$, for O+C,
O+Cu, O+Ag and O+Au collisions at 200 GeV/nucleon (in target
fragmentation region); $(b)$ - the same with $\lambda$ replaced by the
mean number of sources {\bf S}, $\langle C\rangle$ from
eq.(\ref{eq:LL}). Data are from \cite{CAMP,CAMP2}.
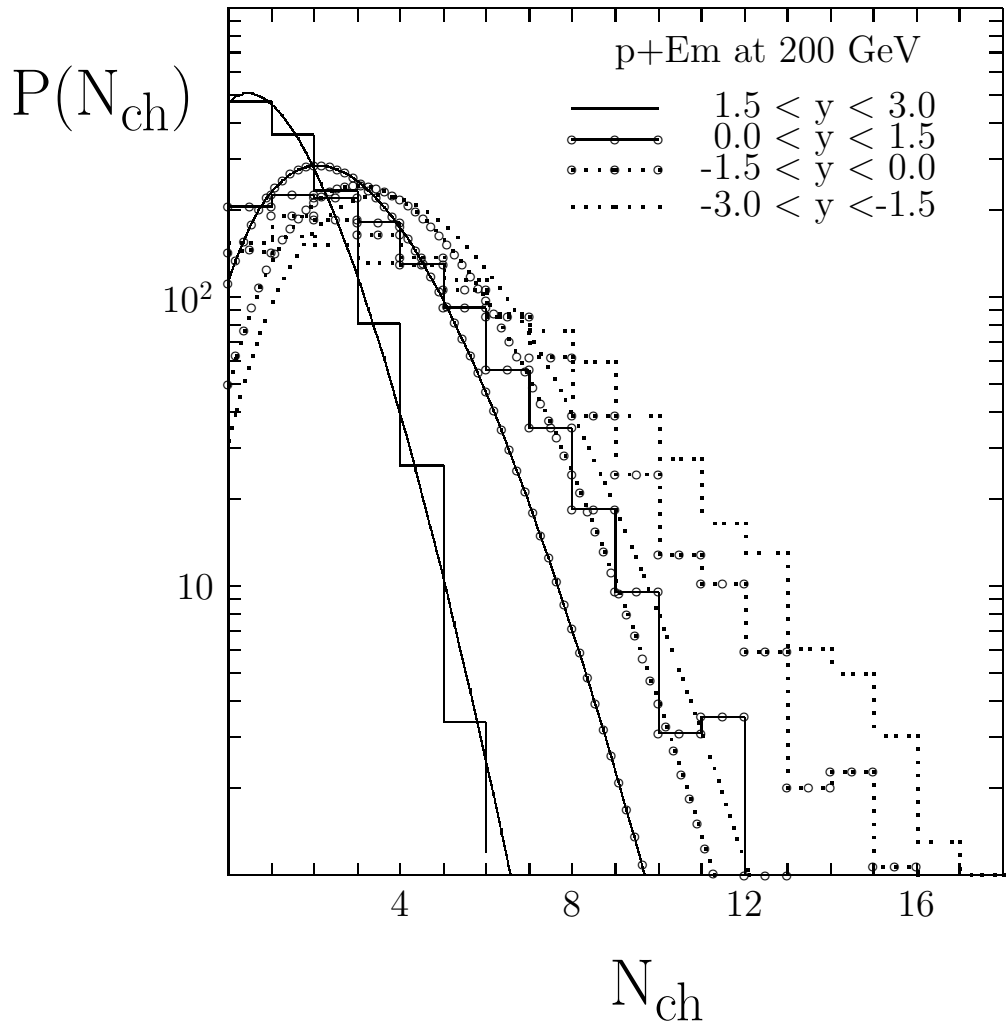
\item[{\bf Fig. 5}] Multiplicity distributions (for p+Em at
$\sqrt{s}=20$ GeV) for rapidity windows covering different quarters of
the kinematically accepted range $-3 < y < 3$ compared with Poisson
distributions with the same value of $\langle N_{ch}\rangle$.
\end{itemize}

\newpage
\begin{figure}
\setlength{\unitlength}{0.240900pt}
\ifx\plotpoint\undefined\newsavebox{\plotpoint}\fi
\sbox{\plotpoint}{\rule[-0.200pt]{0.400pt}{0.400pt}}%


 \end{figure}
\vspace{3cm}
\begin{center}
{\huge Fig. 5}
\end{center}

\newpage
\begin{thebibliography}{99}
\bibitem{MOD} There is a whole spectrum of such models ranging from
statistical, quantum statistical, cascade, to branching or string ones,
to name but few, where sources are called 'fireballs', 'firetubs',
'clusters', 'clans', 'droplets' etc. Cf., for example, R.Hagedorn, {\sl
Nuovo Cim. Suppl.} {\bf 3} (1965) 147; G.N.Fowler et al., {\sl Phys.
Rev. Lett.} {\bf 57} (1986) 2119; G.Espong, Proc. $16^{th}$ Int. Symp.
on Multip. Dynamics, Kiryat Anavim 1985, ed. J.Grunhaus, Editions
Frontiers, Gif-sur-Yvette, 1985, p. 309;  K.Werner, {\sl Phys. Rev.
Lett.} {\bf 61} (1988) 1050.
\bibitem{B} A.Giovannini and L.Van Hove, {\sl Z. Phys.} {\bf C30}
(1986) 391.
\bibitem{ee} R.Szwed, G.Wrochna and A.K.Wr\'oblewski, {\sl Mod. Phys.
Lett.} {\bf A6} (1991) 245.
\bibitem{pp} R.Szwed, G.Wrochna and A.K.Wr\'oblewski, {\sl Mod. Phys.
Lett.} {\bf A6} (1991) 981.
\bibitem{AA} A.Bamberger et al., (NA35 Collab.), {\sl Phys. Lett.} {\bf
B205} (1988) 583.
\bibitem{CD} J.A.Casado and S.Date, {\sl Phys. Lett.} {\bf B344} (1995)
441.
\bibitem{LOR} B.L\"orstad, {\sl Int. J. Mod.Phys.} {\bf A4} (1989)
2861; D.H.Boal, C.-K.Gelbke and B.K.Jennings, {\sl Rev. Mod.Phys.} {\bf
62} (1990) 553.
\bibitem{DEUT} M.Deutschmann et al., {\sl Nucl. Phys.} {\bf B204}
(1982) 333. Cf. also T.Peitzmann, {\sl Z. Phys.} {\bf C55} (1992) 485
and {\bf C59} (1993) 127.
\bibitem{WEINER} G.N.Fowler and R.M.Weiner, {\sl Phys. Rev.}
{\bf D14}, 3118 (1978); M.Biyajima, {\sl Prog. Theor. Phys.} {\bf 66}
(1981) 1378. Cf. also R.M.Weiner, {\it A Strategy for Multiparticle
Dynamics}, invited talk at the Ringberg Workshop on Multiparticle
Production, Fluctuations and Fractal Structure, eds. R.C. Hwa, W. Ochs,
N. Schmitz,  Ringberg Castle, Germany, June 25-28, 1991, World
Scientific 1992, p. 321-339.
\bibitem{GKW} M.Gyulassy, S.K.Kauffmann and L.W.Wilson, {\sl Phys.
Rev.} {\bf C20} (1979) 2267.
\bibitem{QQ1} R.Lednicky, V.L.Lyuboshits and M.I.Podgoretskii, {\sl
Sov. J. Nucl. Phys.} {\bf 38} (1983) 147.
\bibitem{BOWA} B.Andersson and W.Hofmann, {\sl Phys. Lett.} {\bf B169}
(1986) 364; M.G.Bowler, {\sl Phys. Lett.} {\bf B185} (1987) 205 and
{\bf B276} (1992) 237; X.Artru and M.G.Bowler, {\sl Z. Phys.} {\bf C37}
(1988) 293.
\bibitem{BMW} M. Biyajima, T. Mizoguchi, and G. Wilk, {\sl Z. Phys.}
{\bf C65} (1995) 511.
\bibitem{F1} Cf., for example, data presented in \cite{CAMP} and in
\cite{DATA} for ISR, SPS and Tevatron energies.
\bibitem{CAMP} Proc. on {\it Int. Workshop on Correl. and Multip. Prod.}
(CAMP), Marburg, FRG, May 14-16, 1990, eds. M.Pl\"umer et al., World
Scientific 1991.
\bibitem{DATA} A.Breakstone et al. (ISR-SFM Collab.), {\sl Z. Phys.}
{\bf C33} (1987) 333; C.Albajar et al. (UA1 Collab.), {\sl Phys. Lett.}
{\bf B226} (1989) 410; T.Alexopoulos et al., (E735 Collab.), {\sl Phys.
Rev.} {D48} (1993) 984.
\bibitem{CAMP1} J.Pluta et al. in \cite{CAMP}, p. 49.
\bibitem{CAMP2} T.Peitzmann et al. (WA80 Collab.) in \cite{CAMP}, p. 107.
\bibitem{CC} P.Carruthers and C.C.Shih, {\sl Int. J. Mod. Phys.} {\bf
A2} (1987) 1447.
\bibitem{BAR} S.Barshay, {\sl Nucl. Phys.} {\bf B238} (1984) 227.
\bibitem{SR} P.P.Srivastava, {\sl Phys. Lett.} {\bf B198} (1987) 531.
\bibitem{F2} Comparison with data on $P(n)$ performed with NB
distributions suggests \cite{B} that $\langle C\rangle $ increases
with energy from $5$ to $\sim 8$ and saturates at this value for
energies above that of ISR. Another example of such superposition is
$P(n|\sqrt{s}) = \int ^1_0 dK \chi(K) P(n|W(K))$
where $P(n|W(K))$ is the multiparticle distribution of the type
discussed above for $e^+e^-$ but corresponding only to the part of the
total energy of the reaction given by $W(K)=K\cdot \sqrt{s}$ with $K$
denoting inelasticity of the reaction under consideration \cite{INEL}.
It is widely known that Poissonian $P(n|W)$ with $<n(W)>$ being random
variable obeying gamma distribution (e.g., because inelasticity
fluctuations described by the inelasticity distribution $\chi(K)$)
leads again to the NB type of multiplicity distribution (with $k$ being
parameter of gamma distribution this time \cite{CC}).
\bibitem{INEL} G.N.Fowler, R.M.Weiner and G.Wilk, {\sl Phys. Rev.
Lett.} {\bf 55}, 173 (1985); G.N.Fowler et al. {\sl Phys. Rev.} {\bf D35},
870 (1987) and {\sl Phys. Rev.} {\bf C40}, 1219 (1989).
\bibitem{ZL} Zhuang Pengfeu and Liu Lianshou, {\sl Phys. Rev.} {\bf
D42} (1990) 848.
\bibitem{CCS} C.C.Shih, {\sl Phys. Rev.} {\bf D34}
(1986) 2710.
\bibitem{MUP} I.Derado, G.Jancso and N.Schimitz, {\sl Z.Phys.} {\bf
C56} (1992) 553.
\bibitem{NA35} J.B\"achler et al. (NA35 Collab.), {\sl Z.Phys.} {\bf
C57} (1993) 541.
\bibitem{F} K.Fia\l kowski, {\sl Z.Phys.} {\bf C61} (1994) 313.
\bibitem{AG} Depending on our taking as an elementary source
either the whole $e^+e^-$ process or each of its two quark jets. In
fact, when energy increases and gluonic jets start to show up they can
be counted as additional sources in the second case, cf., for example,
R.Ugoccioni, A.Giovannini and S.Lupia, {\sl Z.Phys.} {\bf C64} (1994)
453 and references therein.
\bibitem{IGM} G.Wilk, R.M.Weiner and Z.W\l odarczyk,  {\it The Monte
Carlo Event Generator for Modelling the Coherence and Chaos in High
Energy Hadronic and Nuclear Collisions}, Report SINS - 2153/PVIII/1993;
cf. also: Proc. of the {\sl Workshop on Pre-equilibrium Parton
Dynamics in Heavy Ion Collisions}, LBL, August 23 - Sept. 3, 1993; ed.
X.N.Wang, LBL-Report 34831 (1993).
\bibitem{FFW} G.N.Fowler, E.M.Friedlander and R.M.Weiner, {\sl Phys.
Lett.} {\bf B104} (1981) 239.
\bibitem{ZW} Z. W\l odarczyk, Proc. of the {\sl 23th Int. Cosmic Ray
Conference}, Calgary, July 10-30, 1993, eds. D.A.Leahy, R.B.Hicks and
D.Venkantesan, World Scientific, Singapore (1994), p. 355;
G.Wilk and Z.W\l odarczyk, {\it Fluctuations and Correlations Beyond
Midrapidity};  to appear in the Proc. of the {\sl 24th
Int. Cosmic Ray Conf.}, Rome, August 28 - Sept. 8, 1995 and in
preparation.
\end{thebibliography}
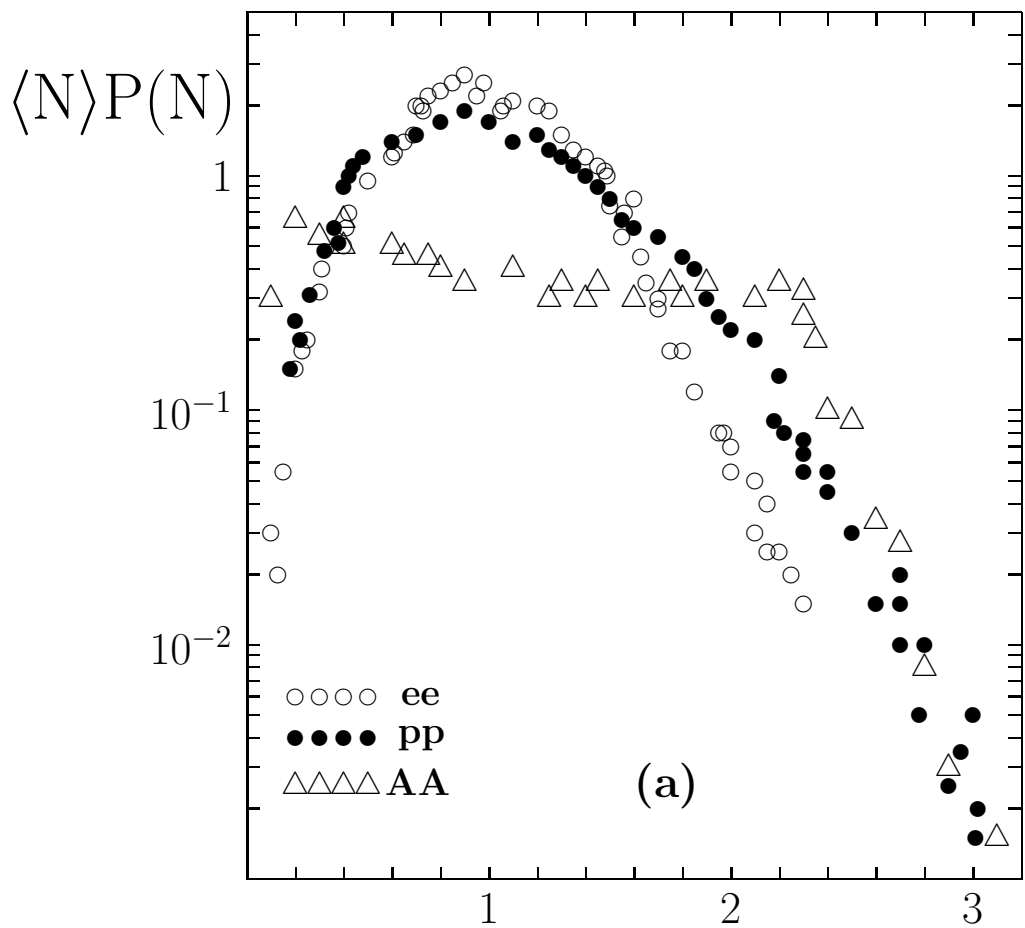
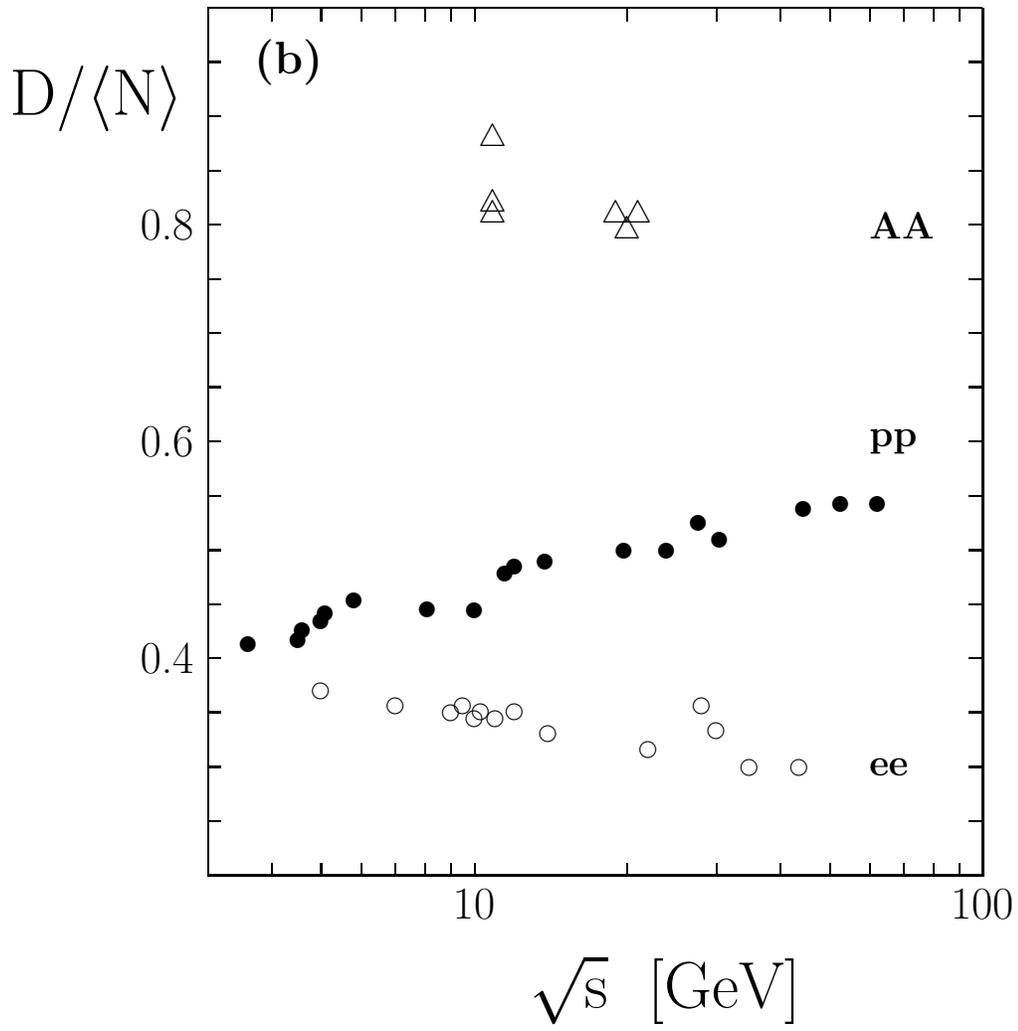
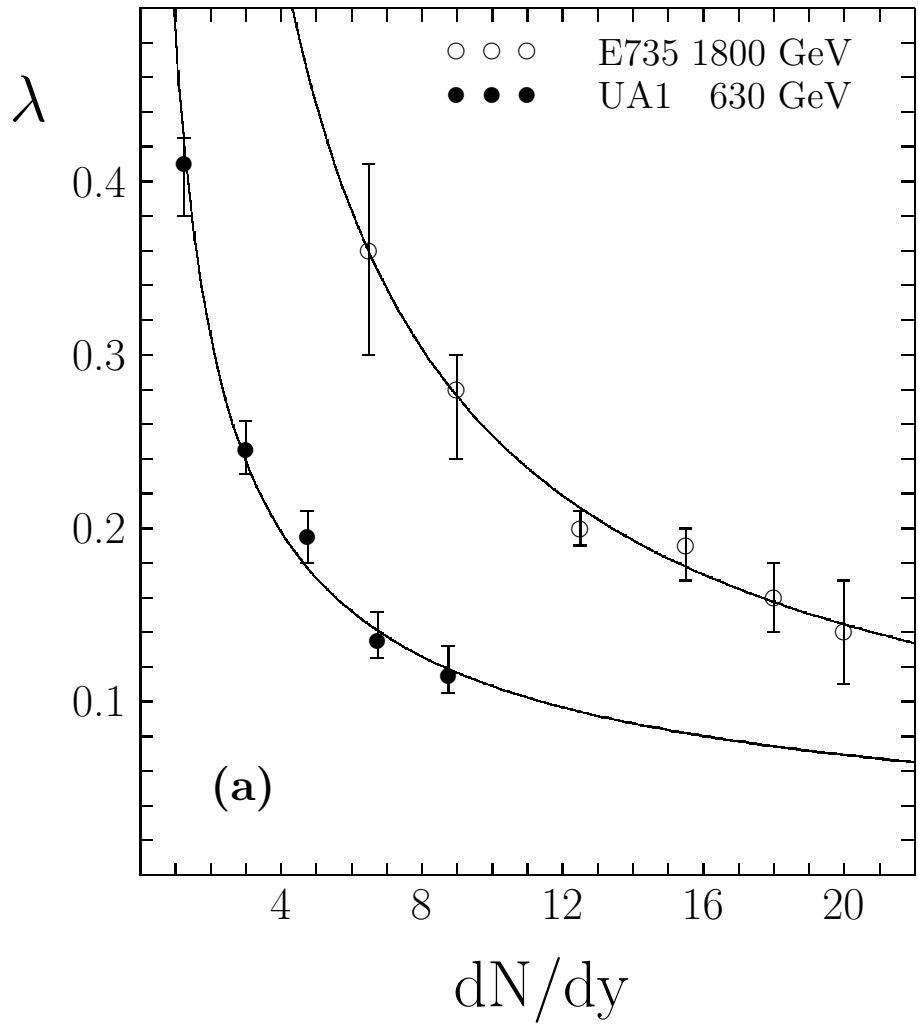
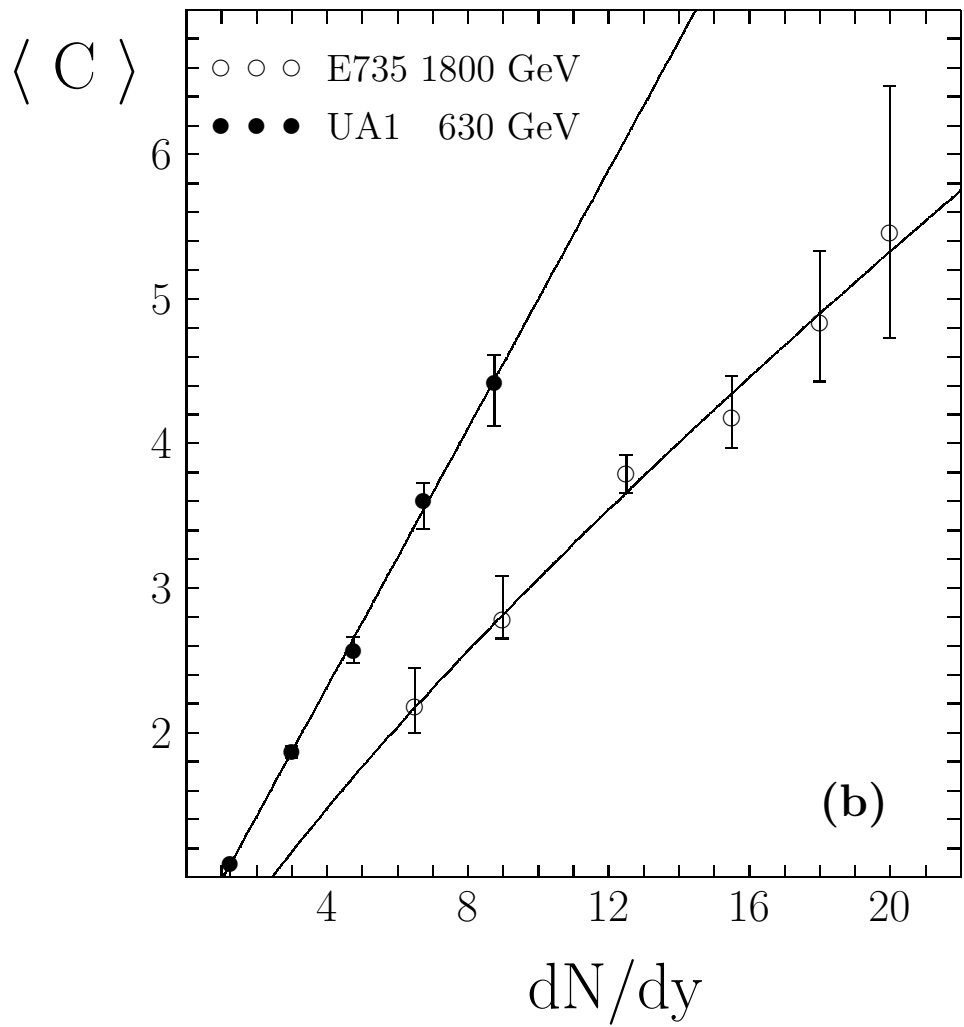
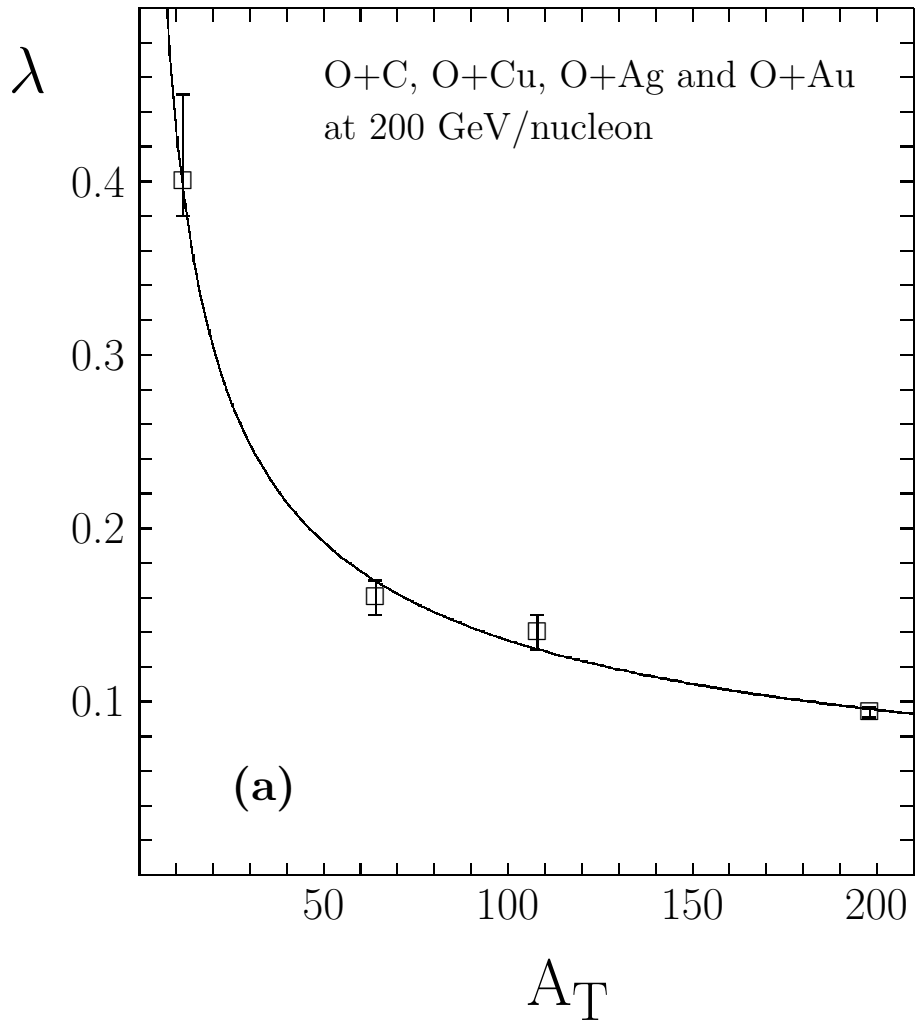
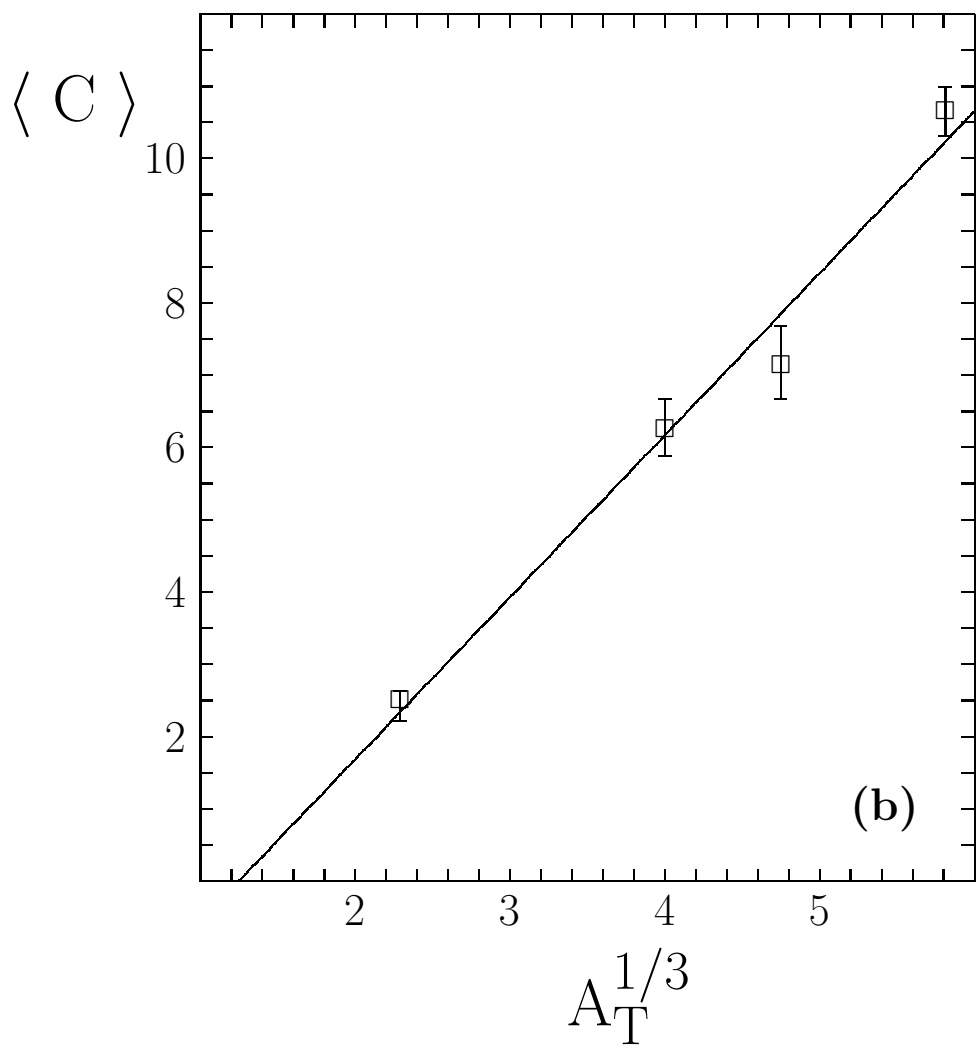
\end{document}